Elizaveta Sivak[1*] and Ivan Smirnov[1]

[1] Institute of Education, National Research University Higher School of Economics, Myasnitskaya ul., 20, Moscow 101000, Russia
*esivak@hse.ru

# Measuring adolescents' well-being: correspondence of naïve digital traces to survey data

**Abstract.** Digital traces are often used as a substitute for survey data. However, it is unclear whether and how digital traces actually correspond to the survey-based traits they purport to measure. This paper examines correlations between self-reports and digital trace proxies of depression, anxiety, mood, social integration and sleep among high school students. The study is based on a small but rich multilayer data set ($N$ = 144). The data set contains mood and sleep measures, assessed daily over a 4-month period, along with survey measures at two points in time and information about online activity from VK, the most popular social networking site in Russia. Our analysis indicates that 1) the sentiments expressed in social media posts are correlated with depression; namely, adolescents with more severe symptoms of depression write more negative posts, 2) late-night posting indicates less sleep and poorer sleep quality, and 3) students who were nominated less often as somebody's friend in the survey have fewer friends on VK and their posts receive fewer "likes." However, these correlations are generally weak. These results demonstrate that digital traces can serve as useful supplements to, rather than substitutes for, survey data in studies on adolescents' well-being. These estimates of correlations between survey and digital trace data could provide useful guidelines for future research on the topic.

## 1 Introduction

Adolescents' everyday lives and well-being are a black box to researchers. It is not well-known how adolescents' behavioral patterns, moods, emotions, and other psychological states change over time. Youths are particularly difficult to study due to uncertainty about the accuracy and validity of adolescents' self-reports [1–2]. Also, surveys are too expensive and time-consuming to be conducted frequently. Meanwhile, adolescents' well-being is attracting greater attention from researchers and policy-makers as mental health issues among adolescents increase [3–4].

Digital traces (data from social media and other digital platforms) present a promising new approach to studying adolescents' well-being that is fast, inexpensive, non-intrusive, and with high resolution. Previous literature has demonstrated that mental health conditions, such as depression and anxiety, can be predicted from mobile sensor data [5–8], social media engagement [9], language [9–12], and photos [13]. An increasing number of studies have analyzed emotive trends based on social media data [14–15].

However, further research is needed to better understand how digital traces correspond to survey data and whether they can act as a substitute for surveys. It remains unclear whether and to what extent digital traces correspond to the characteristics they are assumed to reflect such as sleep patterns, social relations and integration, mood, and psychological well-being. High school students have been understudied as previous studies have typically focused on undergraduate students or adults [16–19]. Unfortunately, the results of these studies cannot necessarily be generalized to adolescents since high school students may use social media differently than undergraduates or adults. Thus, it is uncertain whether digital traces from social media can elucidate the psychological states or behavior of high school students.

Moreover, some online indicators of well-being have been studied much less than others. For instance, researchers have mostly used data from wearable devices and mobile sensors to predict sleep patterns [19–21], but these kinds of studies may be difficult to implement on a large scale. Using data that are more readily accessible, such as timestamps on posts or logins to a learning management system [22], could allow social scientists to study sleep patterns in a way that is both fine-grained and large-scale. However, there is still little evidence that such indicators reflect actual sleep patterns.

Another understudied measure of well-being is negative social ties. For adolescents, peer unpopularity and rejection are associated with depression and externalizing problems [23–25]. However, social media sites usually do not collect information about negative links, i.e., "dislikes" [26]. Therefore, it is important to study whether it is possible to infer negative links and unpopularity from the data available on social networking sites.

In this paper, we explore how naïve digital traces correspond to survey data on high school students' well-being, focusing on depression, anxiety, mood, social integration, and sleep. By naïve digital traces we mean digital trace data that seem to reflect behavior, attitudes, or states. The study is based on a unique data set regarding adolescents' social networks, social media use, psychological well-being, sleep, and demographics ($N$ = 144). A public, anonymized version of the data set is available online.[1] Data were obtained from VK, Russia's most popular social networking site, and from adolescents' self-reports on depression, anxiety, and social ties. We also assessed participants' mood and sleep daily over a period of four months using a mobile app. This study analyzes whether the sentiments of adolescents' social media posts are correlated with depression and anxiety and how day-of-week sentiment dy-

1 Due to the sensitive nature of the data, the files are encrypted. The password to decrypt the files will be sent upon request.
https://osf.io/b57rp/?view_only=9ae0229e3cd64452a1ca1573e2b0186d

namics is correlated with self-reported mood. We also investigate how the time of adolescents' posts corresponds to their sleep patterns and quality. Finally, we examine the relationships between (a) interaction-based metrics for online friendship and popularity on social media and (b) actual friendship ties and peer popularity and unpopularity.

# 2 Data and Methods

## 2.1 Participants

This study was conducted among students from one high school in Moscow, Russia. Participation was voluntary. All students between the ages of 16 and 17 were informed about the study and the opportunity to participate. The researchers held group meetings with all students who self-selected to participate where we explained the aims of the study and what data would be collected. Each participant who decided to participate, as well as one of his or her parents, signed an informed consent form. Students were informed that their participation was voluntary and that they could stop participating in the study at any time and request that any data already collected be deleted (to date no one has requested this). The baseline survey was filled out by 144 students, the endline by 78 students, and 118 participants answered everyday questions about mood and sleep at least once.

## 2.2 Procedures

The study took place over a 4-month period (November 2017–February 2018). Participants were asked to fill out a survey at the beginning and end of this period. Both surveys included items that measure depression and anxiety. In the baseline survey, we also asked about friendship and peer popularity and unpopularity; in the endline survey, we asked about sleep quality. Participants were also asked about their current mood three times per day via a mobile app, using the experience sampling method. Similarly, participants were asked once per day what time they had woken up that morning and gone to bed the previous night. Public data were gathered from profiles on VK with the participants' consent. All procedures used to obtain data are described in Table 1.

## 2.3 Measures

**Depression**

We used the Patient Health Questionnaire scale (PHQ-9)[2], which is used to calculate the severity of depressive symptoms. It includes nine items scored from 0 to 3 and generates a severity score of 0 to 27. Scores of 5, 10, 15, and 20 represent the cut-off points for mild, moderate, moderately severe, and severe depression, respectively

2 https://www.phqscreeners.com/

[27]. The scale has been shown to be a valid tool in detecting depression among adolescents across various cultures [28–33].

We measured depressive symptoms twice. The test-retest reliability coefficient is 0.73 ($P < 10^{-11}$). In the analysis, we used the results of the first measurement as it has fewer missing values.

According to the PHQ-9 scale, 10% of the sample had no symptoms of depression (scored 0-4), 41% exhibited mild symptoms (5-9), 26% moderate (10-14), 11.5% moderately severe (15-19), and 11.5% severe (20-27). These rates are unusually high: previous studies that used PHQ-9 to measure depression among school students reported the prevalence of moderately severe/severe depression to be 5–9% [28–29, 34–35]. In Russia, there is not enough data on the prevalence of depressive symptoms in adolescents to compare our results to. We attribute this high rate to the fact that the school in this study is selective and students may be under a great deal of academic pressure and stress. There is also a possibility of self-selection bias.

**Anxiety**

To assess anxiety, we used the subscale of Spielberger's State-Trait Anxiety Inventory (STAI) [36] measuring trait anxiety—that is, anxiety as a personal characteristic. The test-retest reliability coefficient for the two measures of anxiety is 0.82 ($P<10^{-14}$).

**Mood**

We assessed mood daily over a period of 4 months via experience sampling using a 5-point Likert scale from 1 (very bad) to 5 (very good). Mood was assessed three times per day at random points within specific time periods: morning (8:00 a.m.–9:00 a.m. on school days, 10:00 a.m.–11:00 a.m. on days off), afternoon (3:00 p.m.–5:00 p.m.), and evening (8:00 p.m.–10:00 p.m.). We then computed the participants' average mood level for each day of the week.

**Sleep**

We assessed participants' sleep using a mobile app. Each morning, the app asked participants what time they had gone to bed the previous night and woke up that morning. We also assessed sleep quality using the Pittsburgh Sleep Quality Index (PSQI) [37].[3] The PSQI is a self-report questionnaire that assesses sleep quality over a 1-month interval. The measure consists of 19 individual items, creating seven components that produce one global score. Each item is weighted on a 0–3 interval scale. The global PSQI score is then calculated by totaling the seven component scores, providing an overall score ranging from 0 to 21, where lower scores denote healthier sleep quality.

[3] https://www.sleep.pitt.edu/instruments/

**Sentiments of Posts**

We analyzed the sentiments of posts written within the 4-month study period ($N$ = 5,371) using the SentiStrength program [38].[4] SentiStrength estimates the strength of positive and negative sentiment in short texts, even for informal language. SentiStrength reports two sentiment strengths: -1 (not negative) to -5 (extremely negative) and 1 (not positive) to 5 (extremely positive). A team of linguists adjusted it for Russian language use.[5] We used the following as measures of the aggregate sentiment of users' posts: (a) average positive and negative sentiment across all posts and (b) proportion of strongly negative (scored 3, 4, or 5 on the scale of negative sentiment) and strongly positive (scored 3, 4, or 5 on the scale of positive sentiment) posts among all user's posts. We also separately computed the average positive sentiment of posts for each day of the week.

**Night Posting**

We computed the proportion of late-night posts (those written between 1:00 a.m. and 5:00 a.m.) to all the posts written over the 4-month study period and also the proportion of days when a participant wrote a late-night post to all the days when a participant wrote a post.

**Online Friendship**

We used several interaction-based measures of online friendship. For each participant, we measured (1) the number of "friends" they had on VK, including (a) those who studied at the same school and (b) the overall number of VK friends and (2) the average number of "likes" per post (a) from students from the same school and (b) overall. For each pair of participants A and B, we determined (1) whether A and B were friends on VK, 2) whether A and B had at least one reciprocal like (from A to B and from B to A), and 3) average intensity of likes per pair (sum of the proportion of B's posts liked by A and the proportion of A's posts liked by B, divided by two).

**Offline Friendship**

In the survey, participants were asked to indicate up to 10 other students from the same school whom they considered friends. Since the sample included only 10% of all students at the school, who were almost all from different classes, we used unilateral friendship nominations as indicators of friendship, i.e., if at least one person named another as a friend, we considered them friends.

**Popularity and Unpopularity**

Participants were asked to name up to 10 of the most popular students, in their opinion, and up to 10 of the most unpopular students at their school. For each participant, we determined if she/he was mentioned as popular or unpopular.

---

[4] http://sentistrength.wlv.ac.uk/

[5] http://sentistrength.wlv.ac.uk/#Non-English

**Table 1.** Measures and data sources

| Data sources | Measures |
| --- | --- |
| Surveys (base-line/endline) | • Sleep quality (PSQI scale), endline<br>• Friendship ties (with whom are you friends?), baseline<br>• Popularity (who do you consider popular/unpopular?), baseline<br>• Depression (PHQ-9 scale), baseline and endline<br>• Anxiety as a trait (trait anxiety subscale of the STAI), baseline and endline<br>• Gender<br>• Grade ($10^{th}$ or $11^{th}$) |
| Daily self-reports (mobile app RealLife Exp) | • Bedtime/wake up time (asked once per day in the morning)<br>• Mood (assessed three times per day using a 5-point Likert scale, from 1 "very bad mood" to 5 "very good mood") |
| Public data gathered from profiles on social networking site (VK.com) | • For each public post:<br>– Timestamp<br>– Number of likes (from students who attend the same school and overall)<br>– Strength of negative and positive sentiment expressed in the post<br>• Number of friends from the same school (based on the list of school students) and overall |

# 3 Results

## 3.1 Psychological Well-Being: Depression, Anxiety, and Sentiments of Posts

We found that severity of depression is correlated with the proportion of strongly negative posts (Pearson's r = 0.24) and the average strength of negative sentiment expressed in posts (Pearson's r = 0.23). Pearson correlation coefficients are provided in Table 2. Negative emotions were, on average, more pronounced in the posts of participants who have symptoms of depression, with an average negative sentiment of 1.33, than in posts of students with no depressive symptoms, who had an average negative sentiment of 1.09 (P = 0.002). Students with moderately severe or severe depression wrote more posts, on average, that were strongly negative (22% of posts) than those with moderate or mild symptoms of depression (8.6% of posts) and wrote ten times more strongly negative posts than adolescents with no signs of depression (2% of posts). However, we found no significant correlation between anxiety and the sentiments of posts (see Table 2).

For sentiment analysis, we selected only participants who had written at least three posts within the study period. Fig. 1 presents a histogram of the distribution of participants' (*N* = 144) number of posts. As shown in Fig. 1, most participants wrote approximately 5–10 posts. There were two outliers who wrote more than 200 posts who are not shown in the figure and were not included in the analysis. Given the distribution, we have chosen the threshold of *n* = 3. Fig. 2 shows how the correlation coefficient between depression score and the mean strength of negative sentiments expressed in posts depends on the sample size for different cutoffs for the number of posts. Note that for most values of *n* the results are not significant due to the small sample size.

**Table 2.** Pearson correlation coefficients between depression, anxiety, and sentiments of posts

| Variable | Pearson's *r* | P | 90% CI |
|---|---|---|---|
| | Depression | | |
| Average strength of negative sentiment (participants with 3–200 posts, *n* = 61) | 0.23 | 0.07 | (-0.02, 0.42) |
| Proportion of strongly negative posts (participants with 3–200 posts, *n* = 61) | 0.24 | 0.06 | (0.03, 0.43) |
| | Anxiety | | |
| Average strength of negative sentiment (participants with 3–200 posts, *n* = 61) | 0.08 | 0.54 | (-0.13, 0.28) |
| Proportion of strongly negative posts (participants with 3–200 posts, *n* = 61) | 0.14 | 0.28 | (-0.07, 0.34) |

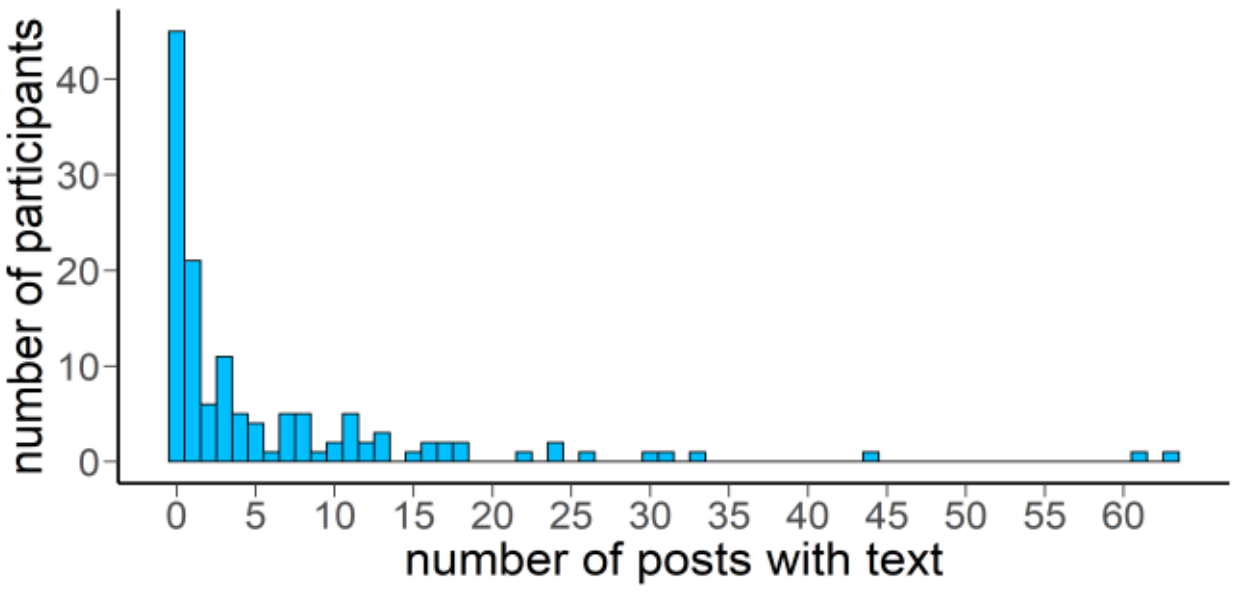

**Fig. 1.** Distribution of participants' number of posts with text

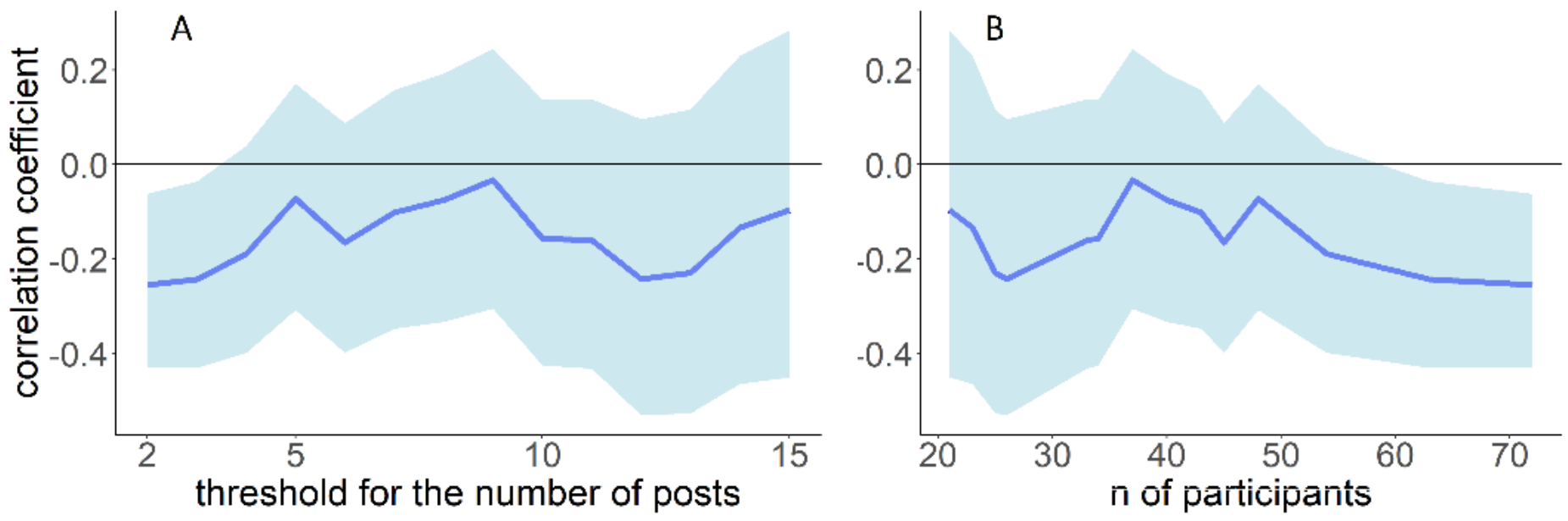

**Fig. 2.** Correlation coefficients between the depression score and the mean strength of negative sentiment expressed in posts with different thresholds for the minimum number of posts (shaded region is 90% confidence interval). The x-axis shows the threshold (Fig.2A) and the corresponding number of participants for each of the thresholds (Fig.2B).

## 3.2 Mood

According to the self-reports acquired via experience sampling, students' positive affect level was higher on weekends and Thursdays, when students leave school for a full day to attend classes of their choice at a partner university, than on school days (see Fig. 3[6] and Table 3). The sentiments of participants' posts on VK in general capture this pattern where the average positive sentiments of posts were higher on weekends and Thursdays than on school days. However, likely due to the small sample size and the lack of statistical power, these estimates did not differ significantly.

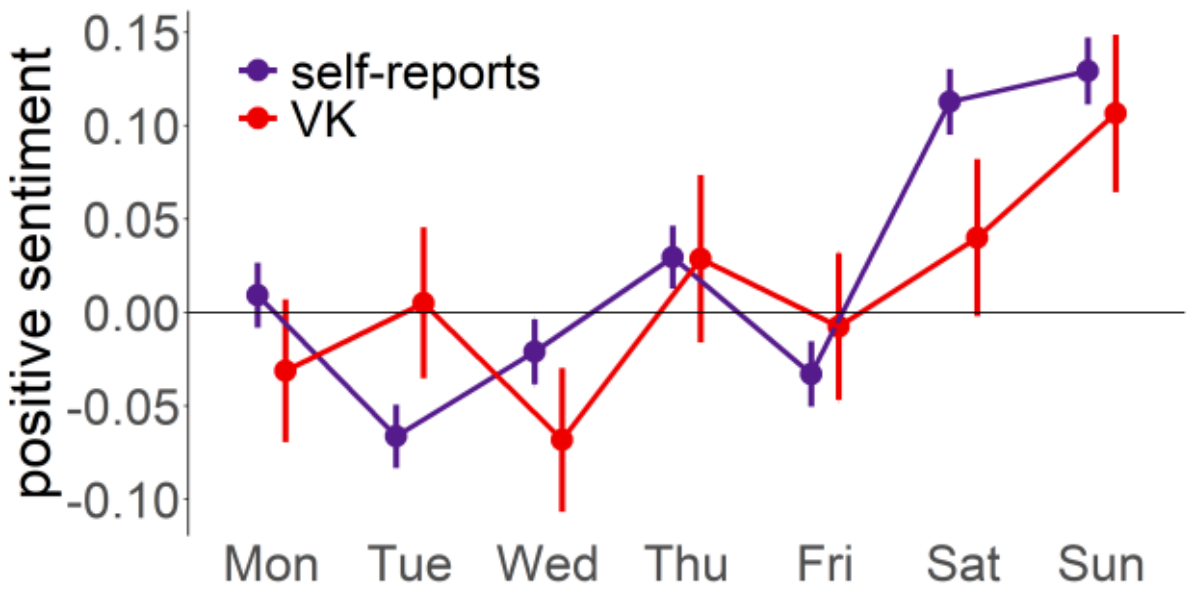

.

**Fig. 3.** Average level of positive sentiment (in SD units) over the course of the week based on self-reports and sentiment analysis of VK posts. Error bars indicate standard errors.

[6] We used a bootstrap approach for hierarchically structured data to compute the standard errors of the mean for positive sentiment and mood on each day of the week as well as 90% confidence intervals for the mean positive sentiment and mood on weekdays and weekends as a correction for repeated measures within the same participant. We implemented a bootstrap approach with replacement on the level of measurements http://biostat.mc.vanderbilt.edu/wiki/Main/HowToBootstrapCorrelatedData.

**Table 3.** Average mood and positive sentiment of VK posts on school days and days off

| | Mean | 90% CI |
|---|---|---|
| Average mood on school days | 3.38 | (3.37, 3.40) |
| Average mood on days off | 3.50 | (3.49, 3.52) |
| Average positive sentiment of VK posts on school days | 1.43 | (1.38, 1.49) |
| Average positive sentiment of VK posts on days off | 1.48 | (1.43, 1.53) |

### 3.3 Online Activity and Sleep

We analyzed whether the times when participants published posts were related to sleep patterns and found that the proportion of late-night posts (written between 1 a.m. and 5 a.m.) is correlated with late bedtime (proportion of days when a participant went to bed after 1 a.m.; $r = 0.2$, $P = 0.024$). Also, see Fig. 4. The proportion of days when a participant wrote a late-night post to all the days with at least one post on VK is correlated with a late bedtime ($r = 0.25$, $P = 0.01$). Among those who have at least one post written between 1 and 5 a.m., 99% on average go to sleep after midnight (71% after 1 a.m.). Those who don't post late are around 1.5 and 2 times less likely to have a late bedtime (69% go to sleep on average after midnight, 36% after 1 a.m.).

More importantly, we found that late-night posting indicated less sleep as well as poorer sleep quality. Students who posted late at night slept 30 minutes less on average than those who did not. On school days, students who posted late at night slept for 6 hours and 40 minutes, in contrast to 7 hours and 10 minutes for students who had no late-night posts ($P = 0.0002$, 95% CI for the difference of means [28, 30]). On days off, students who posted late at night slept for 7 hours and 50 minutes, compared to 8 hours and 20 minutes for students who had no late-night posts ($P = 0.002$, 95% CI = [28, 32]). Sleep quality was correlated with late bedtime ($r = 0.26$, $P = 0.02$, 95% CI [0.008, 0.44]) and on average was worse for those who had at least one late-night post ($P = 0.02$). The means are 0.87 standard deviations apart (6.2 and 7.7 points).

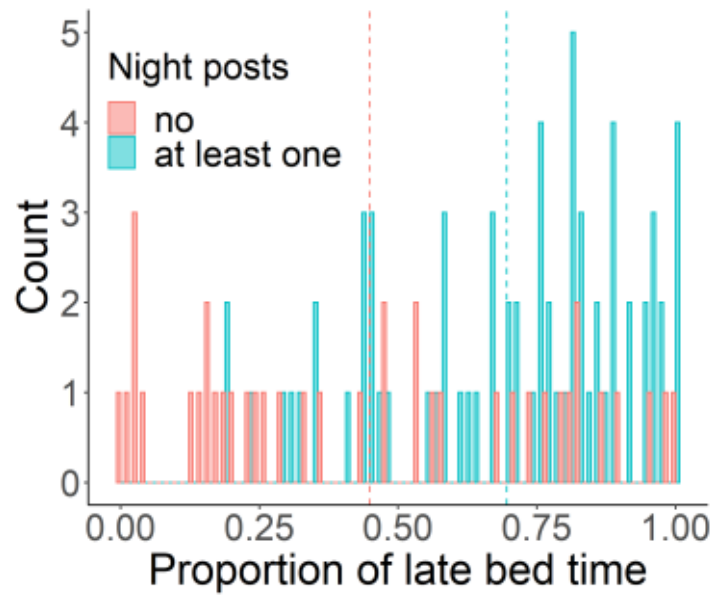

**Fig. 4.** The number of participants with different proportions of late bedtime (days when a participant went to bed after 1 a.m.) for those with at least one late-night post (written between 1 a.m. and 5 a.m.) and those without.

## 3.4 Offline Friendship, Social Status, and Online Behavior

Among all pairs of VK friends, 5.6% are also “offline” friends, according to self-reports. In contrast, the probability of friendship for students who are not friends on VK is only 0.02%.

If a pair of VK friends have at least one mutual “like” (i.e., at least one “like” from A to B and from B to A), the probability that they are “offline” friends is 11%; if they have no mutual likes, the probability is 3.6%. Even for pairs with a high intensity of likes (i.e., every second post is liked), the probability of an offline friendship is only 21%. This indicates that friendship nominations and the intensity of mutual likes reflect different kinds of relationships.

Students who were mentioned as popular had, on average, 1.5 times as many VK friends from the same school than those who were not mentioned as popular (see Table 4). In addition, their posts received more likes on average. Among students in the top 10% by the number of VK friends from the same school, 53% were also named by peers as among the most popular students (top 10% by mentions as popular). Those who were described by peers as unpopular had 1.5 times fewer friends. The same was not true for the overall number of friends on VK, which did not differ significantly for popular and unpopular students. These results suggest that, while online friendship and interactions could be a good indicator of offline social integration, it is important to account for the composition of an online social network, i.e., the overall number of VK friends is not necessarily related to one’s status in a community of peers at school.

**Table 4.** Popularity, unpopularity, and online social ties

| Variable | Named as popular | | Named as unpopular | |
|---|---|---|---|---|
| | Yes | No | Yes | No |
| *n* friends on VK from the same school | 35 | 24*** | 18 | 28*** |
| *n* friends on VK overall | 229 | 196* | 168 | 212 |
| *n* likes from students at the same school | 3.8 | 2.8*** | 2.4 | 3.2*** |
| *n* likes overall | 5.2 | 3.8*** | 2.9 | 4.4*** |

* P = 0.08

*** $P < 10^{-4}$

# 4 Conclusion

Digital traces have the potential to greatly advance research on adolescent well-being. However, ready-made digital traces are framed by the affordances of the service providing the data, its code, cultural usage practices, and users' motives [39]. As such, we should not implicitly assume that digital trace measures accurately reflect adolescents' behavior, attitudes, relationships, or affective states, even if they seem to, as in the case of the sentiments of posts that are often assumed to reflect the author's mood and emotions. Instead, we should test the validity of such naïve digital trace measures among various populations.

In this paper, we analyzed the validity of digital trace data as an indicator of various aspects of high school students' well-being including depression, anxiety, mood, social integration, and sleep. We found that some features of online behavior are correlated with these indicators. For example, the temporal patterns of social media posts corresponded to sleep length and sleep quality. Also, students who were named as unpopular by their peers have, on average, fewer friends on social media at the same school than those who were not unpopular, though unpopular students do not have fewer online friends overall. However, these correlations are generally weak. Even if the absence of significant correlations can be explained by the small sample size, the upper bounds of the confidence intervals indicate no more than a moderate effect. Our results demonstrate that digital trace measures of well-being can be treated and used as complementary data rather than as close proxies.

The estimations of the correlations between digital trace measures and well-being can be useful for potential meta-analyses of the validity of digital trace data as a measure of adolescents' well-being. The data collected here yield novel insights into the relationship between digital markers, social networks, and the well-being of adolescents, considering that this kind of multilayer research on adolescents—combining social media, demographics, depression screeners, and social network data—is rare. The present study can facilitate future research on the validity of digital traces and on the impact of social interactions on adolescents' well-being.